\documentclass[11pt]{article}
\usepackage{Blois_modCE,epsfig}

\def\be{\begin{equation}}
\def\ee{\end{equation}}
\def\bea{\begin{eqnarray}}
\def\eea{\end{eqnarray}}

\begin{document}

%\vspace*{2cm}
%\begin{center}
%\Large{\textbf{XIth International Conference on\\ Elastic and Diffractive Scattering\\ Ch\^{a}teau de Blois, France, May 15 - 20, 2005}}
%\end{center}

% preprint: IFUM-837-FT
% preprint: ECT*-05-17

\vspace*{4cm}
\title{THE ODDERON: THEORETICAL STATUS AND EXPERIMENTAL TESTS}

\author{ C. EWERZ$^{a,b}$}

\address{$^a$ Dipartimento di Fisica, Universit{\`a} di Milano and INFN, Sezione di Milano\\
Via Celoria 16, I-20133 Milano, Italy\\
$^b$ ECT*, Strada delle Tabarelle 286, I-38050 Villazzano (Trento), Italy
}

\maketitle\abstracts{
I review the theoretical and experimental status of the Odderon with 
emphasis on recent developments. 
}

\section{Introduction}
\label{sec:intro}

The Pomeron has a close but less well known relative: the Odderon. 
While the Pomeron is a very interactive chap the Odderon has 
a rather shy personality. That might be a consequence of its 
negative $C$-parity, but is possibly also due to its difficult youth. 
After it was born\,\cite{Lukaszuk:1973nt} in 1973 and obtained its name 
two years later\,\cite{Joynson:1975az} the Odderon experienced a 
hard time in which many of the experts even denied it the right to exist. 
In other words: there was a widely accepted but wrong belief that an 
exchange with negative $C$-parity quantum number persisting 
even at high energies in hadronic collisions could not exist. 
By now the Odderon has become a well-established concept, and 
it is widely appreciated that the Odderon can tell us many 
interesting things about high energy scattering in QCD. 
That is: if we get hold of it -- which turns out to be a surprisingly 
difficult task. 

This year marks the 20th anniversary of the `Blois Workshops 
on Diffractive and Elastic Scattering' as well as the 
30th anniversary of the baptism of the Odderon. 
It is therefore a pleasure for me to review the status of the Odderon 
at this conference in the beautiful surroundings of the Ch\^ateau de Blois. 
The Odderon always attracted considerable interest at the 
Blois workshops, and the proceedings of these conferences offer rich 
material about various aspects of the Odderon and about its ups and 
downs, see especially the talks by Basarab 
Nicolescu\,\cite{Nicolescu:1991ch,Nicolescu:1999qi}. 
In the present contribution I will mainly concentrate on 
recent developments. Ref.\,\cite{Ewerz:2003xi} contains a more 
comprehensive review of the Odderon. 

\section{Basics and Experimental Status}
\label{sec:exp}

The Pomeron carries vacuum quantum numbers and hence is 
even under charge conjugation. The Odderon is also a color singlet 
exchange but has negative $C$-parity and can therefore lead 
to differences between particle-particle and particle-antiparticle 
scattering if it is not strongly suppressed at high energies, that is 
if its intercept is not far below one. In QCD the Odderon can be formed 
by three gluons in a symmetric color state. 

The nonperturbative Pomeron is a complicated but certainly a 
predominantly gluonic object since quark exchanges are in general 
suppressed at high energies. In the simplest picture it can be 
understood as a two-gluon exchange in the $t$-channel. The 
experimentally observed growth of all hadronic cross sections 
with energy provides clear evidence for the Pomeron. Since two-gluon 
exchange can be observed in all hadronic interactions and is 
so important one should expect that three-gluon exchange is at 
least not completely absent. Obviously, one would expect a suppression 
by a power of the coupling $\alpha_s$ for the additional gluon, but for low 
momenta $\alpha_s$ is in fact not too small. According to this 
reasoning there should be a good chance to see (nonperturbative) 
Odderon exchange in a large number of processes. Surprisingly, 
the contrary is true: experimentally the Odderon is extremely difficult 
to find. So far we do not yet have a good explanation 
for this striking result. If we do not find one this might cast 
considerable doubt on our general understanding of high energy 
scattering in terms of gluon exchanges. 

The best but still weak experimental evidence for the Odderon was 
found as a difference between the differential cross sections 
for elastic $pp$ and $p\bar{p}$ scattering at $\sqrt{s}=53\,\mbox{GeV}$ 
at the CERN ISR.\cite{Breakstone:1985pe} 
While the $pp$ cross section has a characteristic 
dip at around $t=-1.3\,\mbox{GeV}^2$, 
the $p\bar{p}$ cross section only levels off at that momentum transfer. 
This difference is typical for a $C=-1$ exchange and cannot be 
explained by mesonic reggeons only. Instead, Regge type models 
with an Odderon as a simple pole\,\cite{Donnachie:iz} 
give a good description of the data, an even somewhat better description 
is obtained with a different Odderon singularity motivated by the 
(functional) maximality of hadronic cross sections.\cite{Gauron:1990cs} 
However, simultaneous data for both $pp$ and $p\bar{p}$ are only available 
for one energy, and are unfortunately not very precise. It seems 
difficult to describe those data without an Odderon contribution, but the 
theoretical description requires to invoke Regge fits with a large number 
of parameters. An important observation is that the Odderon exchange 
is -- in contrast to the Pomeron -- particularly sensitive to the internal 
structure of the colliding hadrons.\cite{Dosch:2002ai} 
Interestingly, both Regge type fits\,\cite{Donnachie:iz,Gauron:1990cs} 
find that the Odderon dominates the region $|t|>3 \,\mbox{GeV}^2$ both 
for $pp$ and for $p\bar{p}$ elastic scattering. Unfortunately, the data in 
this region are only available for a rather small range of energies. Future 
measurements at other energies might offer a chance to study the energy 
dependence of that region and -- possibly -- to identify the Odderon. 
In the future also a measurement of the spin dependence of elastic $pp$ 
scattering at small $|t|$ will offer a good chance to find the 
Odderon.\cite{Leader:1999ua}

Another interesting observable related to Odderon exchange is the difference 
of the $\rho$-parameters for $pp$ and $p\bar{p}$ scattering, where the 
$\rho$-parameter is defined as the ratio of the real and imaginary parts 
of the forward scattering amplitude. Unfortunately, the extraction of the 
$\rho$-parameter from the experimental data is rather difficult, and as 
a consequence the situation regarding the Odderon is not conclusive. 
For a more detailed account we refer to Ref.\,\cite{Ewerz:2003xi}. 

In the observables mentioned so far the Odderon contribution is only one 
of several exchanges in the scattering process and therefore difficult to 
extract. Recently, more attention has been given to exclusive processes 
in which the Odderon is (besides the well-understood photon) the only 
possible exchange. Already the observation of these processes would 
therefore establish the existence of the Odderon, independently of any 
model assumptions. 
An important process of this kind it the diffractive production of 
pseudoscalar and tensor mesons in $ep$ 
scattering.\cite{Schafer:1992pq,Kilian:1998ew} 
While heavy mesons can be produced only at very low rates (see 
Ref.\,\cite{Ewerz:2003xi}) sizable cross sections have been 
estimated for the photoproduction of pions and for $f_2$ and 
$a_2$ mesons\,\cite{Berger:1999ca,Berger:2000wt} 
based on model assumptions for nonperturbative QCD that 
work well in Pomeron exchange processes. So far, however, 
no signal has been observed experimentally. The data for 
pion production lie far below the estimate\,\cite{Adloff:2002dw} 
which can probably be attributed to chiral symmetry effects 
which had not been properly taken into account.\cite{Donnachie:2005bu} 
For the tensor mesons 
only preliminary data are available\,\cite{Olsson:2001nm} 
which are below but not far from the theoretical estimates. 
Effects that could lower the theoretical expectations for the 
tensor mesons (and for the pion) are an unexpectedly low 
Odderon intercept, the suppression of the Odderon-proton 
coupling due to the internal structure of the proton, as well 
as the possibility that the assumptions of the nonperturbative 
model are not adequate to three-gluon correlations relevant 
for the Odderon.\cite{Donnachie:2005bu} Hence, 
although the data do not show an Odderon at the originally 
expected rate they also do not completely exclude it. 

A promising possibility is to look for interference 
effects\,\cite{Brodsky:1999mz} 
between Pomeron and Odderon exchange which is possible in diffractively 
produced final states which are not eigenstates under $C$-parity. 
Charge-asymmetries for example should be sizable (up to 15 \%) 
both in electro-\cite{Hagler:2002nh} and 
photoproduction\,\cite{Ginzburg:2002zd} of pion pairs in $ep$ scattering. 
Other exclusive processes that offer chances to find the Odderon are 
double diffractive vector meson production in hadron-hadron 
collisions\,\cite{Schafer:na}, 
the quasidiffractive process $\gamma \gamma \to \eta_c \eta_c$ at 
a future linear collider\,\cite{Motyka:1998kb,Braunewell:2004pf}, 
and possibly even the diffractive production of exotic hybrid mesons in 
$ep$ scattering.\cite{Anikin:2004ja} 

\section{Recent Theoretical Developments}
\label{sec:recent}

In perturbative QCD the Odderon is described by the BKP 
equation\,\cite{Bartels:1980pe,Kwiecinski:1980wb} which 
generalizes the BFKL equation\,\cite{Kuraev:fs,Balitsky:ic} 
for the Pomeron to three interacting gluons in the $t$-channel. 
The BKP equation exhibits interesting mathematical properties like 
conformal invariance in impact parameter space and holomorphic 
separability\,\cite{Lipatov:1990zb}. 
Furthermore it is equivalent to an integrable system\,\cite{Lipatov:1993yb}, 
namely the XXX Heisenberg chain of SL(2,{\bf C}) spin 
zero.\cite{Lipatov:1994xy,Faddeev:1994zg} 

Two explicit solutions of the BKP equation have been found: 
the JW solution\,\cite{Janik:1998xj} has an intercept of 0.96 
while the BLV solution\,\cite{Bartels:1999yt} has an intercept 
of exactly one which means that its energy dependence is at most 
logarithmic. The wave functions corresponding to the two 
solutions have characteristically different behaviour. While the 
JW solution vanishes when two of the three gluons are at the same 
point in transverse space, the BLV solution is a superposition of 
wave functions in each of which two gluons are always at the 
same point in transverse space. According to this qualitative difference 
of the wave functions the two solutions have very different couplings to 
external particles. 
Here the reggeization of gluons in high energy scattering is crucial, 
which in a nutshell amounts to the fact that whenever two 
gluons are at the same point in transverse space they behave like 
a single gluon. This happens in the BLV solution which is indeed 
a superposition of two-gluon Pomeron eigenfunctions (with odd 
conformal spin to account for the negative $C$-parity). 
In general, one finds\,\cite{Braunewell:2005ct} 
that the most general state of $n$ gluons in the $t$-channel 
can (in leading order) couple only to external particles 
with at least $n$ constituents. If less constituents  are available 
only reggeizing solutions will be projected out of the general wave 
function. Therefore the JW 
solution can couple only to external particles with three constituent, 
for example baryons, but not in leading order to quark-antiquark 
pairs (color dipoles).  
The BLV solution on the other hand can couple to all hadronic particles 
since they always have at least two constituents. 
The different coupling of the two solutions to external 
particles is phenomenologically much more important than the difference 
between their intercepts which for all practical purposes can be 
considered equal. 

There are different approaches to high energy scattering 
in perturbative QCD. For a long time 
the Odderon had been studied only in the traditional 
approach based on the resummation of large logarithms of 
the energy which in the case of the Odderon gives rise to the 
BKP equation. With increasing energy it becomes important 
to take into account also exchanges with more gluons in the 
$t$-channel. In the resummation approach this eventually 
leads to the extended generalized leading logarithmic 
approximation (EGLLA, see Ref.\,\cite{Bartels:1999aw}) 
in which the number of gluons is allowed to fluctuate 
in the $t$-channel. If one considers up to six gluons one can  
obtain a vertex describing the splitting of a Pomeron into 
two Odderons\,\cite{Bartels:1999aw,Bartels:2003zu} from 
a general two-to-six gluon vertex. 
The EGLLA has now also been applied to the Odderon 
channel\,\cite{Braunewell:2005ct} and first steps towards 
the calculation of the effective three-to-five gluon vertex of the EGLLA 
have been made. 

Recently, the Odderon has also been found in other approaches 
to high energy QCD. In the dipole picture the Odderon was found in 
Ref.\,\cite{Kovchegov:2003dm}. In agreement with the argument given 
above only the BLV solution can be obtained in that approach. In the 
color glass condensate approach the Odderon has been discussed in 
the context of classical fields\,\cite{Jeon:2005cf}, and also its quantum 
evolution has been considered in detail.\cite{Hatta:2005as} 
Here both solutions, JW and BLV, can be found when suitable external 
sources are 
considered. Both the color glass condensate approach and the dipole 
picture are particularly well suited for studying the effects of parton 
saturation inside highly energetic hadrons. Interestingly, it turns out 
that the Odderon is suppressed due to saturation if one 
takes into account the effects of Pomeron-Odderon 
interactions in first approximation.\cite{Kovchegov:2003dm,Hatta:2005as} 
Further detailed studies will tell us to what extent this effect can 
explain the apparent smallness of Odderon exchange in many 
reactions that we have described in the previous section. 
Possible phenomenological consequences of Pomeron-Odderon interactions 
have also been discussed in the context of antishadowing in neutrino-nucleus 
collisions.\cite{Brodsky:2004qa} 

From a theoretical point of view only very little is known about the 
Odderon in soft and hence nonperturbative reactions. An interesting 
proposal\,\cite{Kaidalov:1999de} is to infer the intercept of the soft 
Odderon from its Regge trajectory. So far only the masses of the 
lightest two glueballs which can lie on that trajectory, the 
1$^{--}$ and 3$^{--}$, are known and indicate a very low and 
even negative Odderon intercept if a linear trajectory is 
assumed.\cite{Meyer:2004jc} But it is also possible that the 
1$^{--}$ glueball actually lies on a daughter trajectory and then 
the Odderon intercept could be considerably higher and indeed 
close to one. Therefore 
there is a great interest in a future lattice calculation of a 5$^{--}$ 
glueball which would help to clarify the situation of the soft Odderon. 

\section{Summary and Outlook}
\label{sec:summ}

According to our understanding of high energy scattering based on 
the picture of gluon exchanges the existence of the Odderon is very likely. 
Surprisingly though, the experimental evidence for it is rather weak. 
Especially exclusive reactions that can be caused only by the Odderon 
offer good chances to finally establish its existence. 
In perturbative QCD the Odderon is under rather good control. It is 
already by itself a very interesting object from a theoretical point of 
view. It further is an important ingredient in effective theories of 
high energy scattering that are currently discussed. 
New insight into the behaviour of the nonperturbative Odderon 
can be expected from lattice studies of glueball trajectories. 

\section*{Acknowledgments}
I would like to thank the organizers for inviting me to give this talk 
and for creating an inspiring atmosphere. 
This work was supported by a Feodor Lynen fellowship of the
Alexander von Humboldt Foundation.

\end{document}